# Piercing soft solids:
# A mechanical theory for needle insertion

Stefano Fregonese [a], Mattia Bacca [a*]

[a]Mechanical Engineering Department, University of British Columbia, Vancouver BC V6T1Z4, Canada

[*]Corresponding author. *E-mail address*: mbacca@mech.ubc.ca

**Abstract**

In this paper we investigate the mechanical problem of piercing a soft solid body with a needle. This phenomenon is controlled by the critical condition of needle insertion. Needle insertion involves physical and geometrical nonlinearities and a complex failure mechanism. To overcome the complexity of the problem, we describe needle insertion as a sharp transition between two needle-specimen configurations, namely '*indentation'* and '*penetration'*. The sharp configurational change emerges from a mechanical instability and follows the principle of energy minimum. We describe the needle-specimen system in terms of the force applied to the back of the needle and the axial displacement of the needle tip toward the material. At small needle displacements, the energetically favoured configuration is *indentation*. Conversely, when the needle displaces beyond a critical threshold, it penetrates the specimen by rupturing its surface. This creates a new energetically favoured configuration: *penetration*. Our analysis considers a cylindrical needle with a spherical tip, neglects friction and adhesion between the needle and the material, and assumes quasi-static conditions. Despite the mathematical simplicity of our analysis, our theoretical predictions on the needle insertion force have been validated against experiments with surprising accuracy. Our method provides an effective predictive tool, which can be extended to account for different indenter geometry and material behaviour.

*Keywords*: *Puncture; Piercing; Cutting; Soft Materials*

## Introduction

Biomedical operations such as surgery, drug injection [1], biopsy [2], and blood sampling are achieving growing importance in modern medicine especially when automated [3]. However, their success and safety depend on our ability to predict the behavior of biological tissue when rupturing under deep indentation. This is also crucial for understanding the biomechanics of soft tissue injury. In addition to medical applications, a precise knowledge of the mechanisms underlying the phenomenon of cutting, piercing is important for material characterization [4,5], ballistic protection [6], manufacturing [7], and food processing [8].

Throughout evolution, all animal species have evolved with the ability to pierce through and break down biological tissue to feed and defend. The morphology of beaks [9], claws [10], nails

[11], quills [12], and teeth [13] has ultimately reached remarkable geometrical and mechanical properties to ensure success in cutting or piercing with the intended precision. Cutting and piercing, however, have been mainly explored empirically and a comprehensive theory to mathematically describe these processes is currently missing. This is due to the complexity of the mechanical problem, which involves large deformations, large strains, and a complex fracture mechanism. Current models for nonlinear elastic fracture mechanics provided solutions for traditional uniaxial and pure-shear stress states [14]. However, a material being cut is subject to a more complex stress state, not described in previous fracture mechanics models.

In this paper, we focus on the mechanical problem of piercing of a soft solid body with a needle, *i.e.* puncture. The first experimental investigations on the mechanics of puncture were performed on rubber [15], followed by the first theoretical investigations [16]. The latter involved the calculation of the critical force required to deeply penetrate a soft material with needles having a flat or conical tip after the needle is already inserted. This force depends on the toughness of the material, its stiffness, and the radius of the needle. The same authors validated their theory by piercing rubber and porcine skin with needles having various geometries and sizes [17]. However, they did not calculate the critical force required for needle insertion, before deep penetration begins. Previous experiments on rubber [15,17-19], biological tissue [20], and silicone gel [21] evidenced a force peak at needle insertion, followed by a force drop once the needle had pierced through the surface of the specimen and deeply penetrated it.

To unravel the determinants of needle insertion, [22] performed an experimental investigation using needles of various geometries and sizes, puncturing gels with various stiffness and toughness. They described the relation between puncture force $F_c$ and needle radius $R$ with two regimes. For small needle radii, they observed the energy-limited correlation $F_c \sim 2RG_c$, with $G_c$ the toughness of the material. However, $G_c$ measured from puncture experiments is significantly larger than that measured with traditional fracture tests. For larger needle radii, they observed the stress-limited correlation $F_c \sim \sigma_c R^2$, with $\sigma_c$ a 'cohesive stress'. This mechanical property, however, is not measured with other experiments for comparison. Hence, a physical model capable of predicting the condition of needle insertion from material parameters and needle geometry is currently missing.

To overcome the abovementioned limitations, we propose a simple mechanical theory based on a minimum energy principle considering two needle-specimen configurations: *indentation* and *penetration*. Before needle insertion, the needle simply indents the material and the energetic cost associated with it only depends on elastic deformation. At needle insertion, the needle pierces through the material and deeply penetrates it. Needle insertion occurs when *penetration* suddenly becomes the energetically favored mechanism over *indentation*. Our theory is detailed in the next section. We will then provide some numerical results and compare our model against experiments.

**Mechanical theory**

Our theory is based on an energetic comparison between two distinct needle-specimen configurations, namely (*i*) *indentation* and (*p*) *penetration*. Figure 1*a* provides a sketch of configuration (*i*), while Figure 1*b-c* provides a sketch of (*p*). The transition between the two configurations occurs at needle insertion. The needle has a cylindrical stalk with cross-sectional radius $R$ and a spherical tip having the same radius. The axis of the needle is orthogonal to the free surface of the specimen; $d$ is the displacement of the needle tip toward the specimen and $F$ is the force applied to the back of the needle, aligned with the displacement $d$. We consider the specimen to be significantly larger than $R$, so that the specimen size can be ignored. The force applied to the back of the needle is given by the function $F = F(d/R)$ with $F(0) = 0$. The mechanical work done by the force $F$ to push the needle to the depth $d$ is

$$w = \int_0^d F \delta d \tag{1}$$

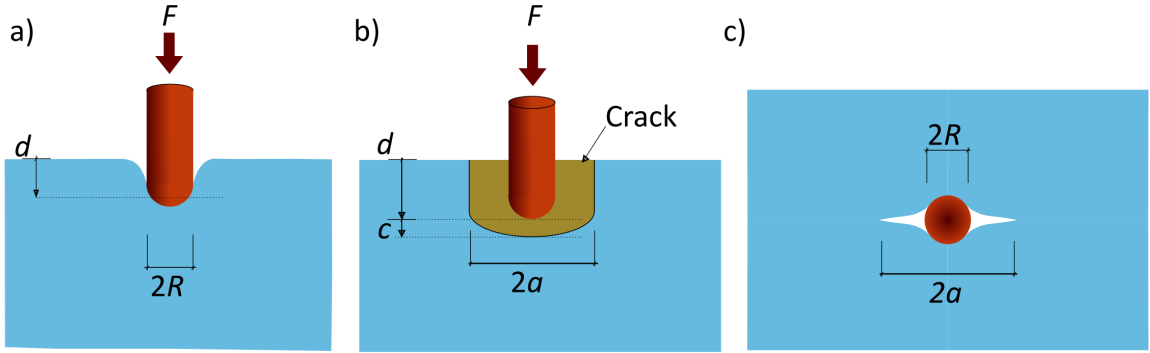

**Figure 1**: Sketch of the two needle-specimen configurations: *indentation* (*i*) *a*) and *penetration* (*p*) *b-c*) (side view and top view, respectively).

The forces $F_i$ and $F_p$ are applied to the needle in the configurations (*i*) and (*p*), respectively. By replacing these forces in Eq. (1) we obtain the mechanical work associated with the two configurations as $w_i$ and $w_p$. The evolution of $F_i$, $F_p$, $w_i$, and $w_p$ for an incrementing depth $d$ is sketched in Figure 2. As evidenced in this figure, needle insertion occurs when $d$ reaches the critical depth $d_c$, at which $w_i = w_p$. For $d \leq d_c$ we have $w_i \leq w_p$, hence (*i*) is the energetically favoured configuration (Figure 2*b*). Conversely, for $d \geq d_c$ we have $w_i \geq w_p$, hence (*p*) is the energetically favoured configuration (Figure 2*b*). The critical force for needle insertion is the force in configuration (*i*) at $d = d_c$, hence $F_c = F_i(d_c)$ as evidenced in Figure 2*a*. This force corresponds to the slope of the curve $w_i$ versus $d$ in Figure 2*b* at $d = d_c$. Once the needle is inserted, the penetration force required to push the needle further down into the specimen is $F_p$ and is equivalent to the slope of the curve $w_p$ versus $d$ in Figure 2*b*. $F_p$ is constant since we neglect friction and adhesion between the specimen and the needle.

We describe the mechanical behaviour of the material with an incompressible single-term Ogden strain energy density functional [23]

$$\psi = \frac{2\mu}{\alpha^2}(\lambda_1{}^\alpha + \lambda_2{}^\alpha + \lambda_3{}^\alpha - 3) \tag{2}$$

In this equation, $\mu$ is the shear modulus of the material (with Young modulus $E = 3\mu$); $\alpha$ is a dimensionless material parameter indicating the tendency of the material to strain-harden; and $\lambda_1, \lambda_2$ and $\lambda_3$ are the principal stretches. The symmetry of Eq. (2) with respect to the three principal directions 1, 2 and 3 is associated with the isotropic behaviour of the material. For $\alpha = 2$, Eq. (2) gives the neo-Hookean form. For larger $\alpha$ the model adopted becomes more representative of rubbers and biological materials, which exhibit a typical *J*-shaped force-displacement curve under uniaxial tensile test. Most biological materials however can exhibit anisotropic behavior; hence Eq. (2) might become unsuitable when a strongly anisotropic behavior is observed. For all other cases, Eq. (2) provides a good generalization of cases by choosing the proper value of $\mu$ and $\alpha$. The Cauchy (true) stress in the material in direction 1 is

$$\sigma_1 = \lambda_1 \frac{\partial \psi}{\partial \lambda_1} - p \qquad (3)$$

with $p$ the hydrostatic pressure applied to the material, *i.e.* a Lagrange multiplier enforcing incompressibility. Eq. (3) can be rewritten for directions 2 and 3 in the same way thanks to the isotropic behaviour of the material.

The needle-specimen system can be described by the dimensionless displacement $d/R$ and the dimensionless force $F/\mu R^2$. Their relation cannot be obtained analytically for either (*i*) and (*p*) configurations. We, therefore, adopt a numerical approach based on finite element analysis (FEA).

*Indentation configuration (i)*

The relationship between $d/R$ and $F_i/\mu R^2$ is calculated via FEA, as detailed in this section. This relation can be generally represented with a series of power-law terms as

$$\frac{F_i}{\mu R^2} = \sum_{j=1}^{\infty} B_j \left(\frac{d}{R}\right)^{\beta_j} \qquad (4)$$

with $B_j$ and $\beta_j$ the power-law-series coefficients, functions of the variable $\alpha$. As reported in *Appendix A*, the results from FEA can be fitted with Eq. (4) with just one power law-term. In this case, we have $B_j = B$ and $\beta_j = \beta$, which values are reported in Table 1 for various $\alpha$.

| $\alpha$ | $B$ | $\beta$ |
|---|---|---|
| 2 | 4.701 | 1.036 |
| 3 | 4.315 | 1.244 |
| 5 | 4.391 | 1.415 |
| 9 | 4.479 | 1.641 |

**Table 1**: Coefficients in Eq. (4) and (5) for various values of the material parameter $\alpha$, obtained via finite element analysis with maximum indentation depth $d/R = 4$.

For very small indentation depths, $d/R \ll 1$ and the Hertzian relationship for linear elastic spherical indentation should apply, giving $F_i/\mu R^2 \approx 16/3 \, (d/R)^{3/2}$ [24]. This would require Eq. (4) to match this solution at $d/R \ll 1$. However, when $d/R < 1$ the surface of the spherical tip is not entirely touching the material while experiments showed that needle insertion occurs at $d_c/R \gg 1$ [22], when the entire surface of the tip is in contact with the material. The same authors also observed that the influence of tip geometry on the material response to indentation reduces significantly at large indentation depths (*i.e.* at $d/R \gg 1$), hence giving more generality to our theory. By substituting the force at Eq. (4) into Eq. (1), we obtain the dimensionless mechanical work

$$\frac{w_i}{\mu R^3} \simeq \frac{B}{\beta+1} \left(\frac{d}{R}\right)^{\beta+1} \quad (5)$$

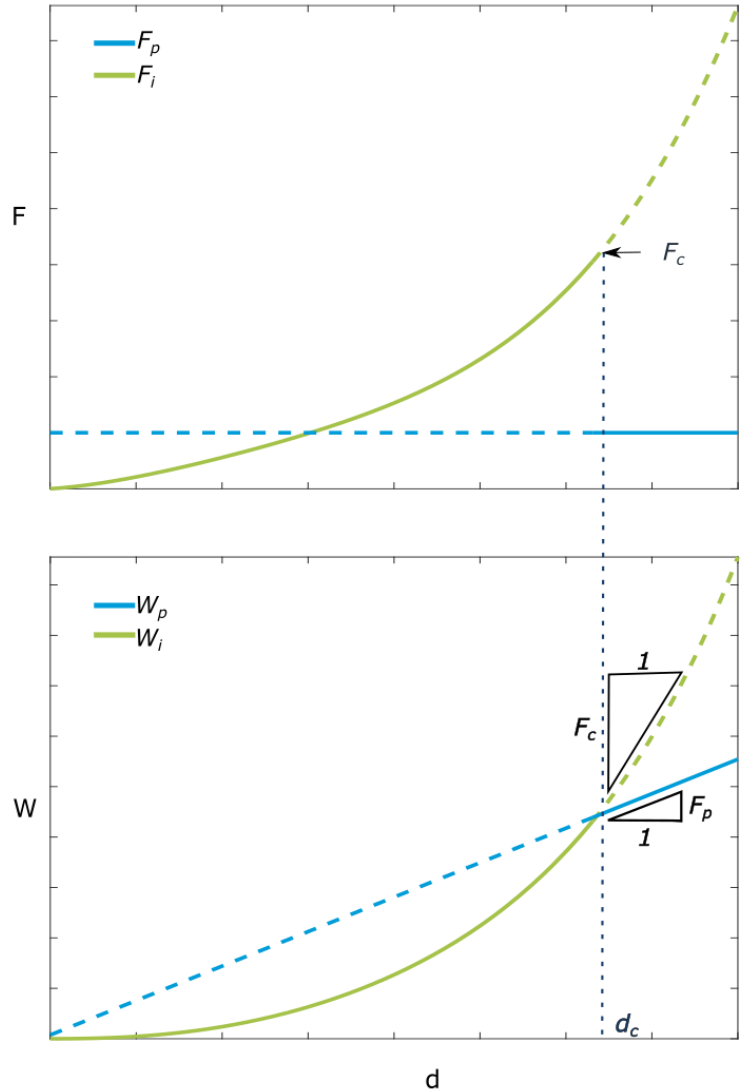

**Figure 2**: Schematic plots of the force $F$ applied to the needle to push it at a depth $d$ into the specimen (top), and the mechanical work $w$ required associated with it (bottom). The blue lines are associated with configuration (*i*) (Figure 1*a*), while the orange lines are associated with configuration (*p*) (Figure 1*b-c*). The solid lines represent the minimum energy path followed by the needle-specimen system.

The coefficients $B$ and $\beta$ in Eq. (4) are obtained from FEA under quasi-static conditions using the commercial software ANSYS. Given the radial symmetry of the problem and the incompressibility of the material, we used the radially symmetric planar elements 182 based on hybrid formulation (see ANSYS' manual for more details). Figure 3*a* sketches the boundary conditions used in the model, with external boundaries fixed. The cylindrical specimen has radius $R_s$ and height $H_s$. To neglect the influence of the specimen size in our results, we tested various values of $R_s$ and $H_s$. For a maximum depth $d_{max} = 4R$, we concluded that a specimen size of $R_s = H_s = 50\,R = 12.5\,d_{max}$ was sufficient to remove its dependency on the results with 1% accuracy. Figure 3*b* shows the finite element mesh adopted. We observed a significant gradient in the strain energy density $U$ of the material in the proximity of the contact region between needle and specimen, as shown in Figure 3*c*. Due to this observation, we adopted a finer mesh near the contact region and a coarser one in the remote regions, near the fixed boundary. The contact between the indenter and the specimen is considered frictionless.

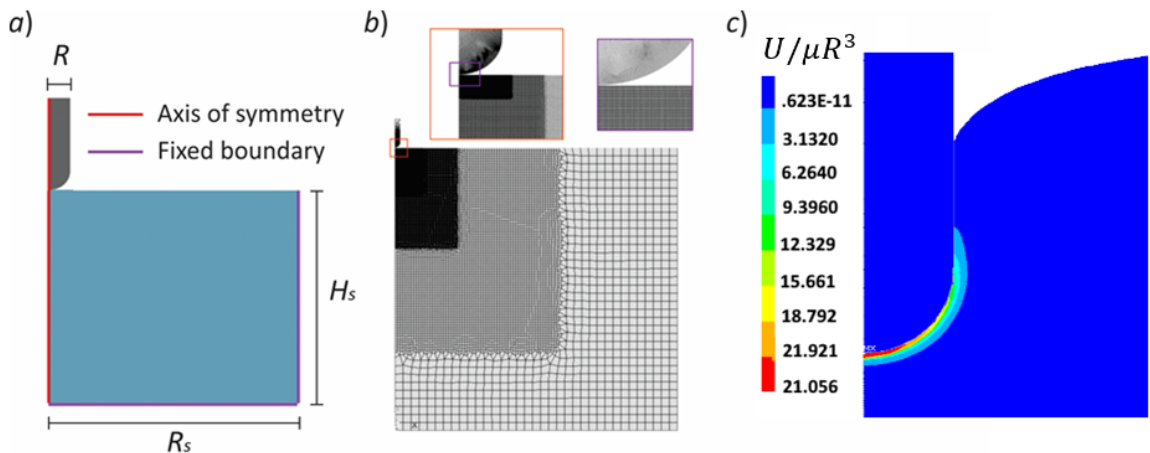

**Figure 3**: Sketch of the finite element model used to describe configuration (*i*): geometry and boundary conditions *a*), mesh structure *b*), strain energy density $U$ distribution *c*).

*Penetration configuration (p)*

The mechanical work $w_p$ required for the needle to penetrate through the material is composed of two main contributions, the work of fracture $w_f$ and the work of spacing $w_s$. The former is the energetic cost required to nucleate and propagate a crack underneath the indenter while the latter is the energetic cost required to 'space out' the material so that the needle can slide into the crack. Both energetic contributions depend on the mechanism of needle penetration. [16-17] observed two failure mechanisms, based on the shape of the indenter. Conical indenters create a planar crack, which propagates in Mode I and is parallel to the axis of the needle. Flat punch indenters instead rupture the material underneath by creating a ring-shaped crack, which propagates in Mode II. This occurs thanks to a shear stress concentration at the perimeter of the contact region, typical of flat indenters. In some cases, however, [25] observed that flat indenters can penetrate the specimen through a planar crack as described for conical indenters. This is because Mode II fracture toughness is commonly much larger than in Mode I.

No specific study unraveled the penetration mechanism of spherical indenters. However, from the observations above, we assume that spherical indenters penetrate the specimen through a planar crack in Mode I, as described in Figure 1b-c.

The work of fracture is given by

$$w_f = G_c \, 2a \, d \, (1 + c/d) \tag{6}$$

where $2a \, d \, (1 + c/d)$ is the area of the crack in Figure 1b-c, and $G_c$ is the toughness of the material. The ratio $c/d$ depends from the properties of the material and scales with the ratio $R/d$. For very soft materials, experimental observations reported large indentation depths at puncture, $d_c$, compared to the radius of the needle [22]. This observation validates the hypothesis of $R/d_c \ll 1$, allowing us to neglect the second term in the parenthesis in Eq. (6).

The work of spacing, considering the penetration mechanism described in Figure 1b-c, is given by the mechanical work required to open the crack so that the needle can slide into it. This can be written as

$$w_s = \mu R^2 \, \hat{h} \, d \tag{7}$$

with $\hat{h}$ a dimensionless parameter, function of the ratio $a/R$. The relation between $\hat{h}$ and $a/R$ is calculated via FEA by [16]. The total work required for the needle to penetrate through the specimen at the depth $d$ is finally calculated by summing the contributions from Eq. (6), with $c/d \approx 0$, and (7). This gives, in its dimensionless form,

$$\frac{w_p}{\mu R^3} = \left(2 \frac{G_c}{\mu R} \frac{a}{R} + \hat{h}\right) \frac{d}{R} \tag{8}$$

By differentiating Eq. (8) with respect to $d/R$ we obtain the dimensionless penetration force

$$\frac{F_p}{\mu R^2} = 2 \frac{G_c}{\mu R} \frac{a}{R} + \hat{h} \tag{9}$$

| $\alpha$ | $a_0/R$ | $C$ | $\gamma$ | $\hat{h}_0$ | $D$ | $H$ | $\delta$ | $\eta$ |
|---|---|---|---|---|---|---|---|---|
| 2 | 0.1603 | 0.1199 | 0.6051 | 0.4478 | 7.8286 | -3.0772 | 0.4612 | 0.3949 |
| 3 | 0.2204 | 2.4904 | 0.6162 | 0.7127 | 3.3553 | -0.9027 | 0.6354 | 0.3838 |
| 5 | 0.4138 | 1.0288 | 0.6088 | 1.0007 | 2.7061 | -0.3676 | 0.7579 | 0.3912 |
| 9 | 0.4237 | 0.6682 | 0.5373 | 1.3154 | 3.08195 | -0.2251 | 0.8603 | 0.4627 |

**Table 2**: Material parameters related to Eq. (10) and (11).

In Eq. (8) and (9), the dimensionless crack size $a/R$ is unknown and should be determined in relation to the material parameters. For a given choice of $\alpha$ and $G_c/\mu R$, the right-hand side of Eq. (9) presents a global minimum in the variable $a/R$, at $a^*/R$. Following the principle of minimum energy, we assume the minimum force at $a = a^*$ to be that at which the needle penetrates the specimen. *I.e.* a small crack can propagate unstably and increase its size $a$ until this reaches $a^*$, after which the crack stops. The relation between $a^*/R$ and $G_c/\mu R$, for various values of $\alpha$, can only be calculated numerically via FEA and has been obtained by [16]. To obtain an explicit relation, we fitted the numerical results to the function

$$\frac{a^*}{R} = \frac{a_0}{R}\left[1 + C\left(\frac{\mu R}{G_c}\right)^\gamma\right] \tag{10}$$

with $a_0$ the nominal crack size, obtained when $\mu R \ll G_c$, and $C$ and $\gamma$ material coefficients that depend on $\alpha$, as reported in Table 2.

Take now $\hat{h}^* = \hat{h}(a^*)$. The relation between $\hat{h}^*$ and $G_c/\mu R$ is also extrapolated numerically from [16], and fitted to the function

$$\hat{h}^* = \hat{h}_0\left[1 + D\left(\frac{\mu R}{G_c}\right)^\delta + H\left(\frac{\mu R}{G_c}\right)^\eta\right] \tag{11}$$

with $\hat{h}_0, D, H, \delta$, and $\eta$ material coefficients given in Table 2 as a function of $\alpha$. All the parameters reported in Table 2 are obtained with the method of the least squares, giving a maximum error comprised within 1%.

By substitution of $a^*$ and $\hat{h}^*$ into Eq. (8) we obtain the mechanical work required for needle penetration $w_p$. Equating the result with $w_p = w_i$, with $w_i$ taken from Eq. (5), we obtain the critical depth $d_c$ as

$$\frac{d_c}{R} \simeq \left[\frac{\beta+1}{B}\left(2\frac{G_c}{\mu R}\frac{a^*}{R} + \hat{h}^*\right)\right]^{1/\beta} \tag{12}$$

with $a^*/R$ and $\hat{h}^*$ taken from Eq. (11) and (12). Substituting this into Eq. (4) we obtain the critical force for needle insertion $F_c$ as

$$\frac{F_c}{\mu R^2} \simeq (\beta+1)\left(2\frac{G_c}{\mu R}\frac{a^*}{R} + \hat{h}^*\right) \tag{13}$$

The dimensionless force required to penetrate the material after needle insertion can be calculated from Eq. (9) by replacing $a$ and $\hat{h}$ with $a^*$ and $\hat{h}^*$, respectively. Comparing this force with Eq. (13), we can conclude that the dimensionless force drop, $\Delta F = F_c - F_p$, produced by needle insertion obeys the simple relation $\Delta F/F_p \simeq \beta$.

**Results**

The critical conditions for needle insertion are described by the dimensionless critical depth $d_c/R$, given by Eq. (12), and the dimensionless critical force $F_c/\mu R^2$, given by Eq. (13). These are functions of the dimensionless parameter $\mu R/G_c$, via Eq. (10) and (11), and the material parameter $\alpha$ (Tables 1-2). Figure 4 reports $d_c/R$ versus $\mu R/G_c$, for various $\alpha$, in a log-log plot, while Figure 5 reports $F_c/\mu R^2$ in the same way. As shown in these figures, $d_c/R$ and $F_c/\mu R^2$ are proportional to the material length scale $G_c/\mu$ and inversely proportional to $R$. $F_c/\mu R^2$ is also proportional to $\alpha$, thanks to a higher strain hardening in the material, and so is $d_c/R$ for $\alpha = 2,3$ and 5. For $\alpha = 9$, $d_c/R$ is smaller than for $\alpha = 5$. This is due to the competition between the mechanical work of indentation, $w_i$ in Eq. (5), and the work of penetration, $w_p$ in Eq. (8). $w_i$ provides the driving force for crack nucleation and propagation, while $w_p$ provides the energetic cost for needle insertion. Both are proportional to $\alpha$, but while an increment in $w_p$ produces an increment in $d_c$, due to higher energetic cost of insertion, an increment in $w_i$ produces a reduction of $d_c$ due to an increased energy accumulation to prompt failure.

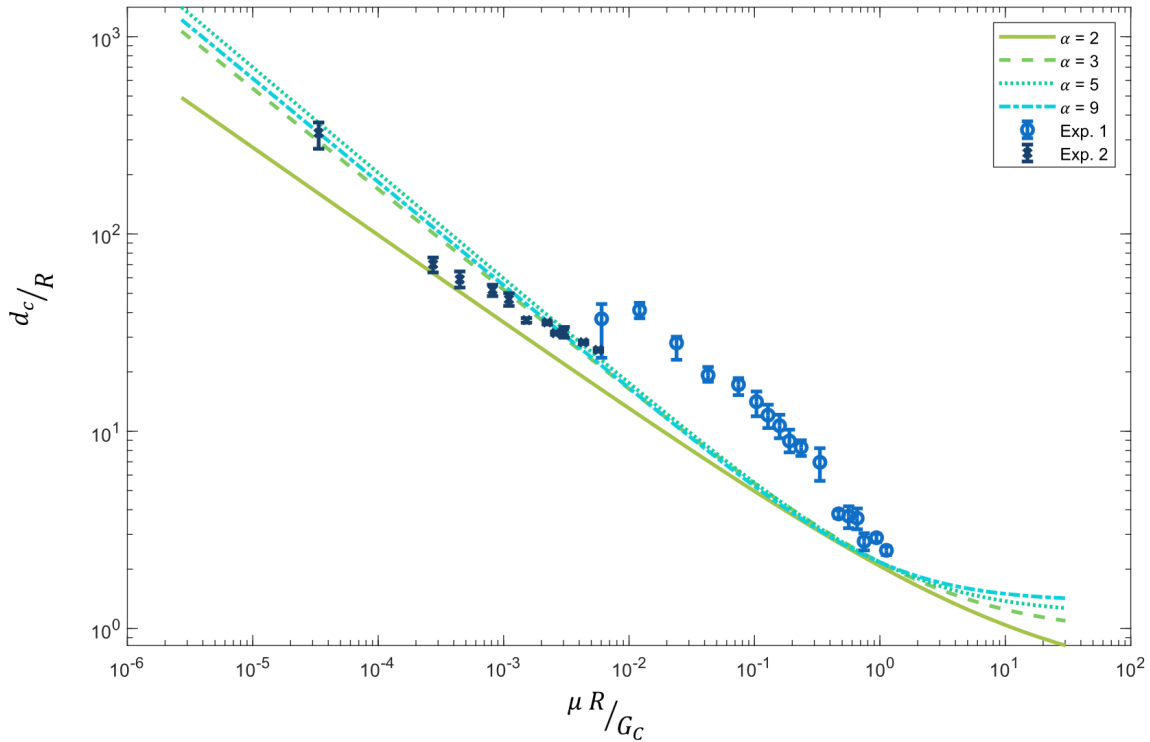

**Figure 4**: Plot of the dimensionless critical depth at needle insertion $d_c/R$ versus $\mu R/G_c$, for various values of $\alpha$, and comparison with experiments (Exp. 1 from [22] and Exp. 2 from [27]).

Figures 4-5 compare our predictions with experiments, taken from [22] (Exp. 1, indicated with circles) and [26] (Exp. 2, indicated with 'x'). These experiments observed the behavior of acrylic triblock copolymer gels punctured with needles having spherical tip of various radii, ranging from $0.1\ mm$ to $5\ mm$ for [22], and from $3.3\ \mu m$ to $70\ \mu m$ for [26]. The shear modulus and toughness of the gels is, respectively, $\mu = 7.2\ kPa$ and $G_c = 25\ J/m^2$ for [22], and $\mu = 54.1\ kPa$ and $G_c = 623.23\ J/m^2$ for [26].

Figure 4 shows a quantitative validation of our model with $\alpha = 9$, if compared against [26], and a qualitative validation against [22]. In the latter case we only capture the trend. In Figure 5 we observe a quantitative validation of our model, with $\alpha = 9$, in comparison with both [22] and [26]. [22,26] only measured the shear modulus of the material, $\mu$, but did not measure $\alpha$. [17] measured $\alpha \simeq 3$ and $\mu \sim 1\ MPa$ with rubbers, and $\alpha \simeq 9$ and $\mu \sim 100\ kPa$ with porcine skin, with the latter having shear modulus in the same order of magnitude as that of the stiffest gel considered here. We adopt $\alpha = 9$ based on the similarity in modulus between porcine skin and the triblock co-polymer gel and validated by the agreement in Figures 4-5. However, a more detailed characterization of the material under indentation may unravel a different value for $\alpha$.

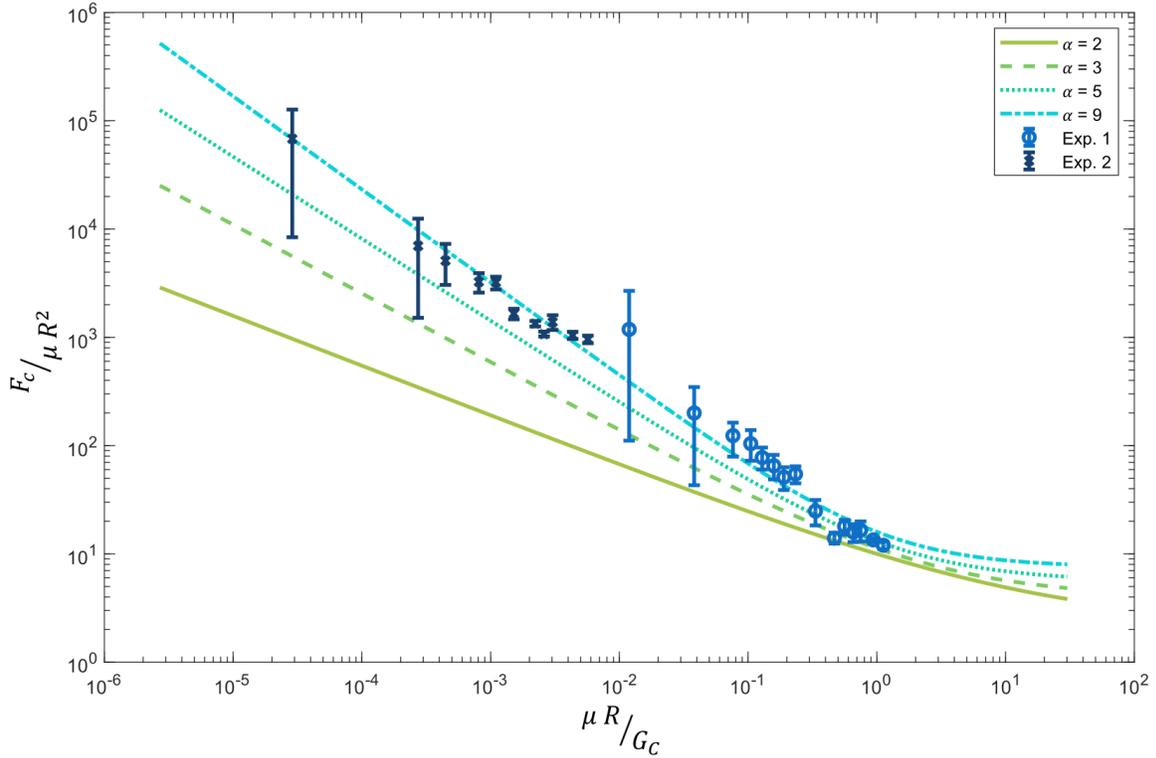

**Figure 5**: Plot of the dimensionless critical force at needle insertion $F_c/\mu R^2$ versus $\mu R/G_c$, for various values of $\alpha$, and comparison with experiments (Exp. 1 from [22] and Exp. 2 from [26]).

**Discussion and Conclusions**

In the proposed theory, the complex mechanism of needle insertion is simply described by a sharp transition between needle indentation and needle penetration, a process driven by *elastic instability*. Our theory is based on *perfect energy transfer*, *i.e.* it relies on the hypothesis that all the strain energy accumulated in the material during indentation, $w_i$, is immediately and entirely available to nucleate and propagate the crack that serves as a channel to accommodate the penetration of the needle. The failure mechanism leading to the formation of the crack-channel is localized underneath the region of contact between the needle and the specimen; hence the strain energy stored by the material via indentation must be accumulated in the same region. As demonstrated in Figure 3*c*, this condition is satisfied.

Our theory neglects the effect of friction and adhesion between the needle and the specimen. During indentation, frictional forces dissipate mechanical energy thereby incrementing the mechanical work needed to obtain needle insertion. Frictional forces between the lateral area of the needle and the specimen during penetration also create energy dissipation. For this reason, $F_p$ is expected to be linearly proportional to $d$, and not a constant as shown in Figure 2 and [16]. This also results in an increment in the energy required for needle insertion and thus an increment in $d_c$ and $F_c$. The quantitative validation shown in Figure 5 suggests that the sensitivity of $F_c$ on adhesion and friction is relatively small, while the mismatch in Figure 4 for [22] suggests that $d_c$ could be significantly affected by interfacial forces in some cases. The mismatch in Figure 4, however, could be attributed to other causes.

Our analysis considers quasi-static conditions. This hypothesis is valid if the time scale of molecular rearrangements in the material is negligible compared to the time scale of needle insertion. *I.e.* the velocity of the needle, $\dot{d}$, satisfies the condition $\dot{d} \ll R/t^*$, with $t^*$ the relaxation time of the material. When this hypothesis is not satisfied, the behavior of the material becomes rate-dependent and the mechanism of needle insertion becomes affected by needle velocity. As demonstrated experimentally [27], an increment in needle velocity produces an increment in $d_c$ and $F_c$. This suggests that viscous forces act as energy dissipators hence requiring additional mechanical work to prompt material failure. A quantitative evaluation of the influence of needle-velocity requires a more sophisticated dynamic FEA, which is left for future development.

The behavior of the material under indentation is analyzed with FE to a maximum depth of $d_{max} = 4\,R$, however in some cases $d_c > 4\,R$. The choice of $d_{max} = 4\,R$ results from a compromise between accuracy and computational efficiency. The behavior of the material under indentation for $d > 4\,R$ is then considered an extrapolation and, as such, can generate inaccuracy. Our prediction on $F_c$ appears to be accurate in spite of this limitation, while the mismatch in $d_c$ for larger radii [22] could be attributed to this.

Our analysis considers needles having spherical tip, however, needle tip geometry can be generalized by proper geometrical assumptions. Experimental observations showed little influence of tip geometry on $d_c$ and $F_c$ for large $d_c/R$ [22].

The simple scaling relation between the needle insertion force and the needle penetration force, $(F_c - F_p)/F_p \simeq \beta$, describes the force drop at needle insertion. This finds qualitative agreement with experiments [17], however, more experimental observations are needed for its validation.

Finally, the mechanical behavior of the punctured material can be further generalized by adopting alternative constitutive models to the one adopted in our analysis. More sophisticated models, however, rely on a large number of parameters, which introduce additional uncertainty and challenges associated with material characterization. The single-term Ogden form in Eq. (2) is an optimal compromise between generality and minimum number of parameters.

Despite the simplicity of the proposed theory and the simplifications above mentioned, we predict the critical force for needle insertion $F_c$ with surprising accuracy.

**Acknowledgments**
We thank David Labonte (Imperial College, London) for helpful conversations. This work was supported by the Department of National Defense (DND) of Canada (CFPMN1-026), by the Natural Sciences and Engineering Research Council of Canada (NSERC) (RGPIN-2017-04464), and by the Human Frontiers in Science Program (RGY0073/2020).

**Appendix A**

Figure A1 reports the plot of dimensionless indentation force, $F/\mu R^2$, versus dimensionless depth, $d/R$. These curves are obtained via finite element analysis as detailed in the *Indentation configuration* Section. As can be seen in the log-log plot of Figure A1 (left), the linear trend suggests a power-law correlation between force and displacement, hence justifying the use of Eq. (4). This is true for $\alpha = 3, 5, 9$ up to the maximum indentation depth analyzed, $d/R = 4$. For $\alpha = 2$, when the material behaves like a neo-Hookean solid, the curve deviates from the simple relation proposed in Eq. (4) for $d/R > 2$. In this case, the simple relation in Eq. (4) is less accurate than for $\alpha > 2$, and a more complicated formulation should be used instead. Because most materials exhibit significant strain hardening, it is common to assume $\alpha > 2$.

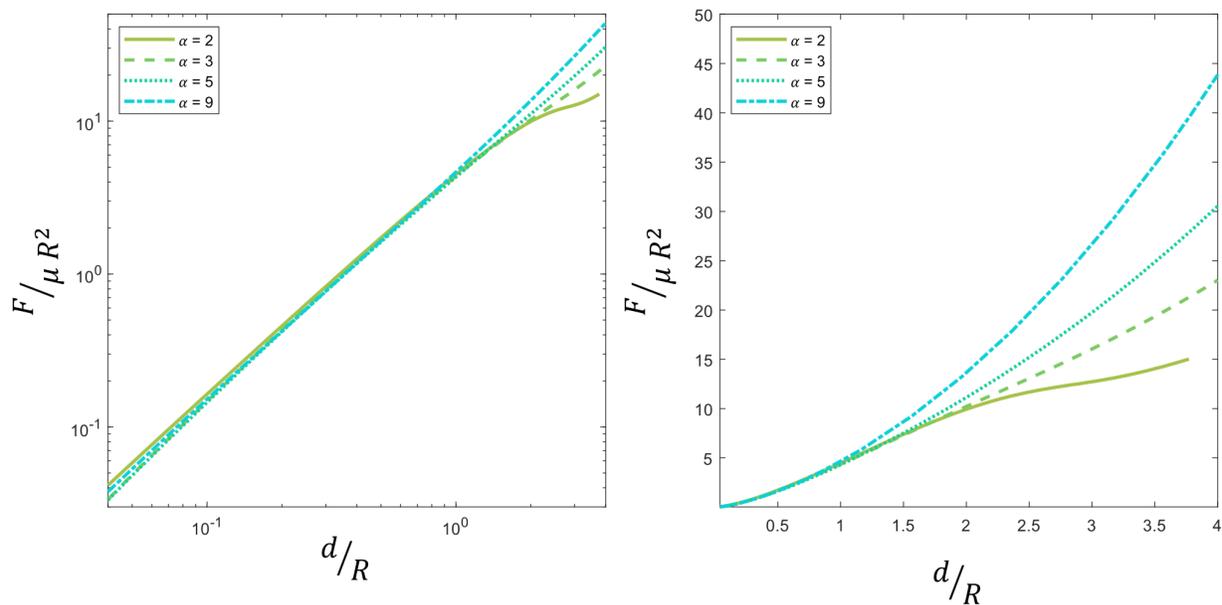

**Figure A1**: Dimensionless indentation force versus dimensionless depth obtained from finite element simulation. On the left is a log-log plot.